\documentclass[%
 aip,
 amsmath,amssymb,
reprint,%
]{revtex4-1}

\usepackage{graphicx}
\usepackage{dcolumn}
\usepackage{bm}

\usepackage[utf8]{inputenc}
\usepackage[T1]{fontenc}
\usepackage{mathptmx}
\usepackage{hyperref}
\begin{document}

\preprint{AIP/123-QED}

\title[Elastic Response of Wire Frame Glasses. I. 2D Model]{Elastic Response of Wire Frame Glasses. I. Two Dimensional Model}

\author{David A. King}
 \email{dak43@cam.ac.uk}
 \affiliation{Cavendish Laboratory, University of Cambridge, J. J. Thomson Ave., Cambridge CB3 0HE, UK}
\author{Masao Doi}%
\affiliation{Centre of Soft Matter and its Applications, Beihang University, Beijing 100191, China
}%
\author{Erika Eiser}
\affiliation{Cavendish Laboratory, University of Cambridge, J. J. Thomson Ave., Cambridge CB3 0HE, UK}%

\date{\today}

\begin{abstract}
We study the elastic response of concentrated suspensions of rigid wire frame particles to a step strain. These particles are constructed from infinitely thin, rigid rods of length $L$. We specifically compare straight rod-like particles to bent and branched wire frames. In dense suspensions the wire frames are frozen in a disordered state by the topological entanglements between their arms. We present a simple, geometric method to find the scaling of the elastic stress with concentration in these glassy systems. We apply this method to a simple 2D model system where a test particle is placed on a plane and constrained by a random distribution of points with number density $\nu$. Two striking differences between wire frame and rod suspensions are found: 1) The linear elasticity per particle for wire frames is very large, scaling like $\nu^2 L^4$, whereas for rods it much smaller and independent of concentration. 2) Rods always \textit{shear thin} but wire frames \textit{shear harden} for densities less than $\sim \sqrt{K/k_B T L^4}$, where $K$ is the bending modulus of the particles. The deformation of wire frames is found to be important even for small strains, with the proportion of deformed particles at a particular strain, $\gamma$, being given by $(\nu L^2)^2 \gamma^2$. Our results agree well with a simple simulation of the 2D system.
\end{abstract}

\maketitle
\section{\label{sec:Intro}Introduction}
Concentrated suspensions of rigid macro-molecules are a prototypical soft matter system, constituting a rich field of study for both experimentalists and theoreticians. The great variety of non-Newtonian flow behaviours displayed by these systems has been a particular interest. Being able to predict and engineer these behaviours based on the shape of the suspended particles is a problem of broad scientific an technological importance.

The interactions between the particles are generally what lead to the pronounced non-Newtonian behaviour. The strongest effects are observed for concentrated suspensions, where the number density of particles, $\rho$, is large such that $\rho L^3 \gg 1$, with $L$ being a typical length-scale of the particles. Hydrodynamic and excluded volume interactions will always be present and have an effect on the rheology, but for certain particle shapes and concentrations a different kind of interaction dominates. These are `kinetic' or `topological' constraints, which originate from the fundamental property that no two particles can cross. This is the case when the particles are very long compared to their width and the particles are rarely in close contact, such that $\rho V_{exc} \ll 1$, where $V_{exc}$ is the excluded volume of the particles. As the excluded volume is irrelevant, \textit{all} equilibrium properties are known; the system is equivalent to an ideal gas. On the other hand, the non-equilibrium properties, such as the rheology, are severely affected by the non-crossing condition. This is often referred to as the `entanglement effect'. 

\begin{figure}\includegraphics[width=8cm]{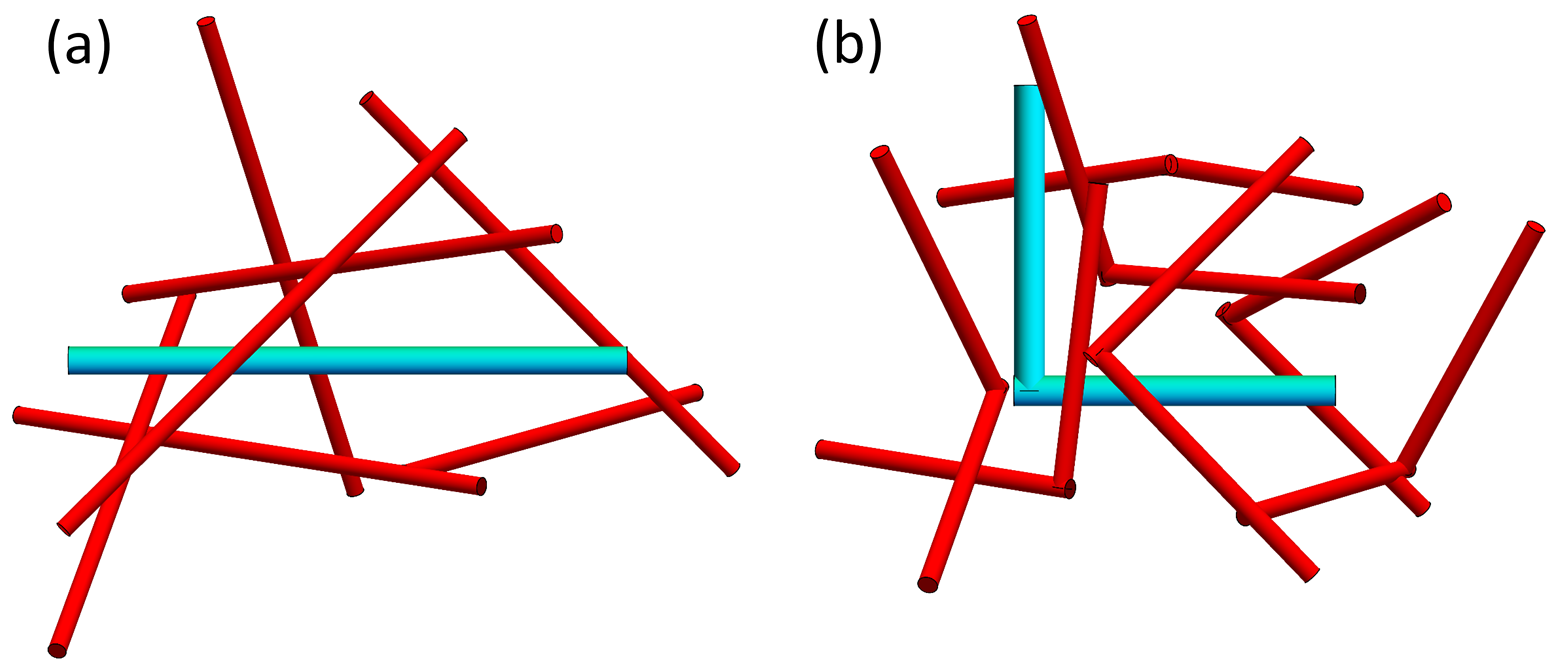}
	\caption{\label{fig:DenseSystems} Sketches of the systems considered. (a) A dense suspension of rod like particles. A test rod is shown in cyan. The motion of the test rod is restricted by the surrounding red rods, but it can still diffuse along its length, so the suspension is still in a fluid state. (b) L-shaped particles in dense suspension. The cyan test L-shape is trapped by the surrounding red L-shapes. If it moves along the length of one of its legs, the other becomes entangled. The system is frozen in a glassy state.}
\end{figure}

We will focus on how kinetic constraints affect the rheology of the suspension depending on the shape of suspended particles. The concentration range considered is, $1/L^3 \ll \rho \ll 1/V_{exc}$. This ensures that the kinetic constraints are the dominant interaction and that there is no long range orientational order in the system. Each of the particle shapes we will consider is constructed from infinitely thin, rigid rods. We consider the joints between the rods to be effectively rigid and unaffected by thermal fluctuations. Such particle shapes can be divided into three different classes; straight, bent and branched. Straight particles are simply rods, see Figs.(\ref{fig:DenseSystems}a) \& (\ref{fig:2DParticles}c). Bent particles are two rods joined at some angle, see Figs.(\ref{fig:DenseSystems}b) \& (\ref{fig:2DParticles}a). Branched particles are those where three or more component rods meet at a point. These can be planar stars as in Fig.(\ref{fig:2DParticles}a) or fully three dimensional shapes, for instance the shapes in Fig.(\ref{fig:2DParticles}d) with an extra rod piercing the plane. We refer to the bent or branched particles as `wire frames'.

Wire frame particles can be realised practically using DNA origami techniques\cite{Seeman1982NucleicLattices, Rothemund2006FoldingPatterns}. Arms made of double stranded DNA sequences can be engineered so that they rigidly link at precise angles to form `nanostars' \cite{Bomboi2019Cold-swappableGels, Biffi2013PhaseNanostars, Xing2018MicrorheologyHydrogels}. Modelling the double stranded DNA arms as infinitely thin, perfectly rigid rods is reasonable, due to their large aspect ratio ($\sim 20$) and stiffness (persistence length $\sim 390$\r{A})\cite{Gross2011QuantifyingTension}. Being able to predict the different rheological responses depending on the shape of the nano-stars is very important for the design of these systems as functional nano-materials.

The kinetic constraints in these systems can also be responsible for a glass transition. Suspensions of 3D crosses, particles consisting of three mutually perpendicular, infinitely thin rods joined at their midpoints, have been shown to have glassy dynamical behaviour in the range of densities we are considering\cite{VanKetel2005StructuralGas}. In this system the translational diffusion coefficient goes to zero with an exponential dependence on the density. The majority of particles are rendered effectively immobile by the kinetic constraints imposed by their surroundings and the system is frozen in an isotropic, disordered state. 

The geometry of the suspended particles must influence this behaviour. In Fig.(\ref{fig:DenseSystems}) we sketch two examples of the systems we are considering. A dense suspension of rods is shown in Fig.(\ref{fig:DenseSystems}a) and a dense glassy state of L-shaped wire frames is shown in Fig.(\ref{fig:DenseSystems}b). In each panel a particular test particle is highlighted in blue. The rod may still diffuse in this concentration regime by the reptation mechanism \cite{Doi1975RotationalSolution, Doi1986TheDynamics}. However, the wire frames cannot. This is because the reptation process relies on the rods' ability to diffuse freely along their length even though their transverse motion is severely hindered. If a wire frame diffuses along the length of one of its constituent rods, then any one of the others will quickly become entangled, blocking its motion. 

It is interesting to compare the rheology of glassy states of wire frames, Fig.(\ref{fig:DenseSystems}b), to dense fluid states of rods, Fig.(\ref{fig:DenseSystems}a). For flexible polymers, it is well known that branched star polymers display significantly different flow behaviour from linear chains, with extremely slow stress relaxation \cite{Doi1986TheDynamics, Gennes1975ReptationJpa-00208365,Pearson1984ViscoelasticPolymers, Doi1980RheologyMelts,Milner1997Parameter-freeMelts}. This is as a result of the branching making reptation impossible. It is reasonable then to expect an equally pronounced difference in behaviour between rigid branched particles and rods. In fact, it has been observed in simulations\cite{Heine2010EffectSuspensions} that the viscosity for dense suspensions of branched 3D crosses is significantly larger than for rods and that the viscosity has a much stronger dependence on concentration \cite{Petersen2010ShearNanoparticles}. 

Evidently, we cannot use the standard approach based on the tube model \cite{Doi1978Dynamics1, Doi1986TheDynamics}, where the reptation mechanism is used in a hydrodynamic formalism. Instead we must use a new approach which determines the rheological properties of these glassy states directly from the constraints on each particle. In this two part series of papers, we restrict our attention to a simple question which gives useful insight into the problem. We do not address in detail the time (or frequency) dependence of the rheological response here, but rather focus on its instantaneous magnitude. Our hope is that the method presented here will form a useful basis for more detailed studies in the future.

We begin with some simple definitions before stating the problem explicitly. The rheological properties of a suspension are expressed by the constitutive equation relating the stress tensor to the applied rate strain tensor, $E(t)$. The stress will consist of two parts; the elastic stress, which we call $\sigma$, and the viscous stress, $\sigma_{V}$. The viscous stress is related to the energy dissipation in the system and can generally be written as a function of time, $\sigma_{V}(t) = \mathcal{V}:E(t)$, where the 4th rank tensor $\mathcal{V}$ can be a function of the deformation history of the suspension. The elastic stress is related to the change in the free energy per unit volume, $F$, by the virtual work principle;
\begin{equation}
\label{virtualwork1}
\delta F =  \sigma : E \delta t,
\end{equation}  
where $\delta F \equiv F(E) - F(0)$ is the change in the free energy, calculated as the difference between the free energy in a system deformed by $E$ and that in the undeformed state. If the applied strain is small enough, the elastic stress will depend on it linearly,
\begin{equation}
\label{elasticresponse}
\sigma(t) = \int_{-\infty}^{t} dt' \ \mathcal{E}(t-t') :E(t'),
\end{equation}
where the time dependent, fourth rank tensor $\mathcal{E}$ is the elastic response function.

To probe the rheology of these systems, we consider the stress response to a step strain. In this case, the rate of strain tensor is given by,
\begin{equation}
\label{stepstrain}
E(t) = \kappa \delta(t),
\end{equation}
where we have defined the strain tensor, $\kappa$. We always take the strain to be simple $x-y$ shear for which $\kappa$ has only one non vanishing component, $\kappa_{xy} \equiv \gamma$. The extension to other strains is straightforward. For step strain, the viscous stress must also be proportional to a delta function, and so can be taken to vanish for all practical timescales. 

The virtual work principle for the elastic stress can be re-written,
\begin{equation}
\label{virtualwork2}
	\sigma = \frac{\partial F}{\partial \kappa},
\end{equation}
and for small strains will take the form,
\begin{equation}
\sigma(t) = \mathcal{E}(t) : \kappa = \gamma \mathcal{E}_{xy}(t),
\end{equation}
hence the choice of step strain allows us to directly probe the elastic response function. 

Typically the stress will start at a particular value, $S_0$, and then decay. The initial timescale for this decay, $\tau_0$, is expected to be the same for all concentrations. This is because immediately after the deformation, the particles are free to move without hindrance from the surroundings. After a time, $\tau_{\text{int}}$, the particles have moved sufficiently to begin interacting with their neighbours and the decay timescale will increase to $\tau(\rho)$, due to the entanglement effect. This new decay rate is a function of the particle density, since for denser systems the entanglement effect is stronger. Therefore a reasonable approximation for the time dependent elastic stress is, 
\begin{equation}
\sigma(t) =
\begin{cases}   
S_0 \ e^{-t/\tau_0} \ , & \text{for} \ t < \tau_{\text{int}} \\
S_0 \ e^{\tau_{\text{int}}(1/\tau(\rho)-1/\tau_0)} e^{-t/\tau(\rho)} \ , & \text{for} \ t > \tau_{\text{int}}
\end{cases}.
\end{equation}
In concentrated suspensions, it is known that $\tau_0 \ll \tau$ and hence we can ignore the initial rapid decay of the stress and take $\sigma(t)$ to be approximated by, 
\begin{equation}
\label{stressdecay}
\sigma(t) \approx S_0 \ e^{-\tau_{\text{int}}/\tau_0} e^{-t/\tau(\rho)} \equiv \sigma_0(\rho) e^{-t/\tau(\rho)},
\end{equation}
where we have defined $\sigma_0(\rho)$ as the initial value of the stress measured in response to a step strain. Throughout these papers, we refer to $\sigma_0(\rho)$ as simply `the elastic stress'.

From this discussion we see that there are two main parts to the stress response. Its initial magnitude, $\sigma_0(\rho)$, and $\tau(\rho)$, the dominant timescale of the subsequent decay. This timescale is generally very long, so the elastic stress persists for a long period after the initial strain. There are many interesting questions relating to this timescale in the case when reptation is not possible. In particular, is there a critical concentration at which this timescale diverges? Or is the relaxation a single or stretched exponential, as could be expected?\cite{Edwards1986TheTransition}. While we do not address these problems here, we hope that the methods we introduce may be of use in their resolution.

In this paper (paper I), we present a simple geometric method for determining the scaling of the elastic stress, $\sigma_0(\rho)$. This method is applicable to a wide range of wire-frame shapes and is outlined in the next section. In section \ref{sec:2dmodel} we introduce a two dimensional model system to which we apply this method. We find a striking difference between straight, rod-like particles and bent or branched particles. In section \ref{sec:nonlinear}, we extend our treatment to the non-linear elasticity and show that there is a critical density, related to the bending modulus of the individual particles, where the behaviour changes from shear hardening to shear softening. This is in contrast to the result for rods, where the suspension always shear thins. Our theoretical results are found to be consistent with those of a simulation of the 2D system. The purpose of this paper is to introduce the method in the 2D model, where the calculation can be performed simply and exactly. A full treatment of the problem in three dimensions will be given in paper II. 
\section{General Method: Linear Response}
\label{generalmethod}
To determine the elastic stress from the virtual work principle (\ref{virtualwork2}), we need the free energy as a function of strain. The free energy is, of course, $F = U - T S$, where $U$ is the internal energy, $S$ is the entropy, and $T$ is the temperature. We consider the system in the absence of an external potential, therefore the strain only changes the internal energy if the particles themselves are deformed. This will be shown to contribute only to the non-linear elastic response, which is dealt with in section \ref{sec:nonlinear}. In this section, we outline a simple, geometric method for calculating $\sigma_0$ from the entropy change per particle caused by the strain in the linear regime.

In the glassy states, the motion of a test particle is impaired by the other particles in the system, with its centre of mass confined to a `cage' and only certain orientations accessible at any given moment. Generally, it is only the Brownian rotations of the particle which contribute to the stress, not the motion of its centre of mass. This allows us to treat the centre of mass as fixed and focus only on the constraints placed on the particle's orientation. 

Let us define the orientation of the particle as $\Gamma$. In two dimensions, $\Gamma$ is represented by the angle, $\phi$, that the particle makes to a reference axis and in three dimensions, three Euler angles specify $\Gamma$. The surrounding particles constrain the test particle to a region in the configuration space. The size of this region depends on $\Gamma$ and the configuration of the surrounding particles $C$, and is written $\Omega(\Gamma,C)$. The test particle will explore all of $\Omega$ over a timescale of roughly $\tau_{int}$, which is generally very short. Therefore the particle essentially instantly equilibrates inside $\Omega$. The entropy associated with this particle then follows from the Boltzmann definition,
\begin{equation}
\label{entropy}
S(\Gamma , C) = k_B \log \Omega(\Gamma , C).
\end{equation}
When the step strain is applied, the surrounding particles will move, and so the accessible volume of configuration space will instantaneously change from $\Omega$ to $\tilde{\Omega}(\kappa)$. The change in entropy caused by this transformation alters the free energy and hence yields an elastic stress in the system. This method is similar to that introduced by Edwards to study flexible polymers with topological constraints and crosslinks\cite{Edwards1967StatisticalI,Edwards1969TheoryMaterial,Edwards_1969}. As the surrounding particles rearrange due to their Brownian motion, the test particle can explore more of the configuration space, so $\Omega$ begins to increase and the stress relaxes. We are interested in the initial value of the elastic stress, $\sigma_0$, so we only need to consider the instantaneous change after the step strain.

If there are $\rho$ particles per unit volume, the free energy per unit volume immediately after the step strain is, 
\begin{equation}
\label{Fenergy1}
F(\kappa) = - k_B T \rho \int d\Gamma \psi(\Gamma) \Big\langle\log \tilde{\Omega}(\Gamma , C \ ; \kappa)\Big\rangle,
\end{equation}
where $\psi(\Gamma)$ is the orientational distribution function, which may be taken to be uniform since the system in a disordered state, and $\langle \cdots \rangle$ denotes averaging over all configurations, $C$, of the constraints. Differentiating this with respect to $\kappa$ will give the stress immediately after the step strain, $\sigma_0$. 

The advantage of equation (\ref{Fenergy1}) for the free energy is that $\tilde{\Omega}$ can be calculated from purely geometric considerations, without detailed knowledge of the dynamics of the particles. In three dimensions the geometry required will become very complicated, especially when defining the constraints placed on the test particle in order to calculate $\tilde{\Omega}$. In two dimensions, however, the geometry is simple enough that the problem may be solved exactly. 
\section{2D Model and Simulation Details}
\label{sec:2dmodel}
The calculation of $F(\kappa)$, becomes quite complex in three dimensions. Therefore, here we first consider a simpler, two dimensional model.

\begin{figure}\includegraphics[width=8cm]{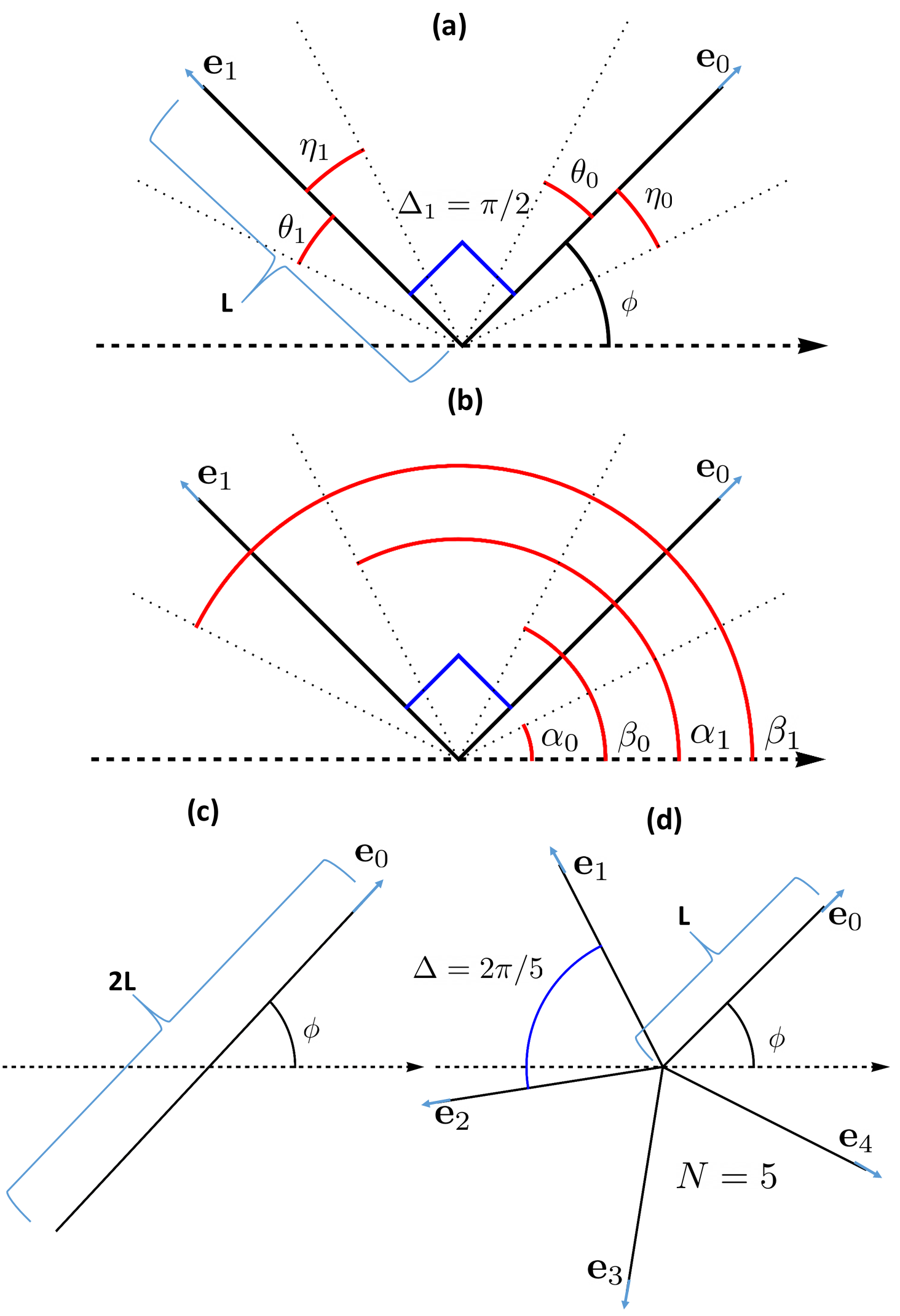}
	\caption{\label{fig:2DParticles} Sketches of the particle shapes considered. (a \& b) An example of a bent particle, an L-shape with arms of length $L$. (c) A rod of length $2L$. (d) An example of a branched particle, a `5-Star'. In each panel the unit direction vectors of the legs, $\textbf{e}_l$, are shown and the orientation of the $l=0$ leg is set at an angle $\phi$ from the $x$-axis, shown as a dashed line. 
	In panels (a) and (b), the L-shape is constrained to lie between the two dotted lines. In (a) the angles, $\eta_{l}$ and $\theta_{l}$, from each leg to the nearest anti- and clockwise constraint respectively, are shown. In (b) the angles which the constraints make to the $x$-axis are shown; $\alpha_l$ for the closest anti-clockwise constraint to leg $l$ and $\beta_l$ for the closest clockwise constraint. The points representing the constraints (see Fig.(\ref{fig:ConstDeform})) are omitted from this sketch for clarity. 
	The angle from the $l=0$ to $l$th leg is $\Delta_l$. Shown in (a), this is $\pi/2$ for the L-shape. For the 5-star in (d) this is $2 l \pi /5$.}
\end{figure}

\begin{figure}\includegraphics[width=8cm]{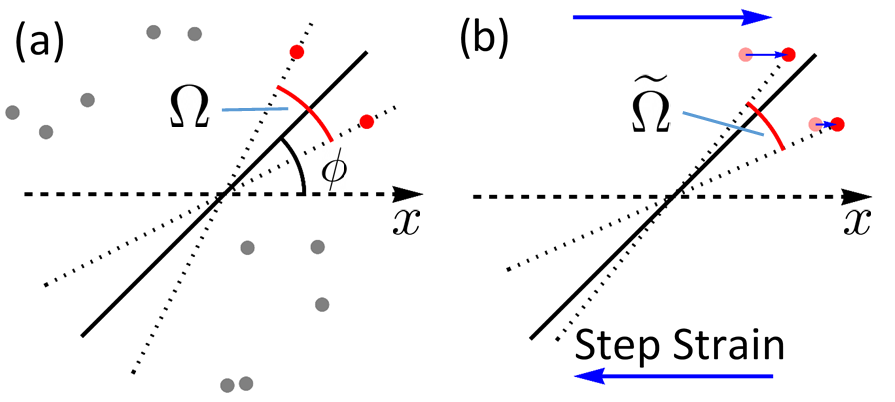}
	\caption{\label{fig:ConstDeform} A sketch of the 2D model system for rods. The rod is shown as a solid line at an angle $\phi$ from the dashed $x$-axis. Panel (a) shows the constraining points surrounding the rod with the two points closest to the rod highlighted in red. These two points constrain the orientation to the range of angles, $\Omega$. In (b) the step strain has been applied, indicated by the blue arrows, and the constraining points closest to the rod have moved, with their new positions shown in the darker, solid colour and their previous positions in a lighter colour. This changes the range of accessible angles to $\tilde{\Omega}$.}
\end{figure}

In this model we only consider planar wire frames, as shown in Fig.(\ref{fig:2DParticles}) and each particle is fixed parallel to the $x$-$y$ plane. The wire frame particles are infinitely thin so no two lie in the same plane and as such, each particle forms its own independent, 2D system. The rotation of each particle in the plane is constrained by many point-like obstacles which it cannot cross, as shown in Fig.(\ref{fig:ConstDeform}a). The number of obstacles per unit area is, $\nu$, and in the concentrated limit, $\nu L^2 \gg 1$. These points play the role of the other wire frame particles in three dimensions, preventing a test wire frame from freely rotating from one orientation to another. We ignore the translational part of the entropy and only consider the rotation of the test particle. The obstacles are also assumed to be fixed and are displaced affinely when the deformation is applied. This model system can be thought of as a cross section of a three dimensional system taken in the plane of the test particle. The obstacle points are then the surrounding particles which intersect this plane.

The particles we consider are all formed of rigid rods of the same length, $L$, referred to as `legs', which will be indexed by $l$. Each leg has the same length, $L$, and the legs all meet at the same point which is fixed at the origin. The direction of the $l$th leg is given by the unit vector, $\textbf{e}_l$, running parallel to it and pointing away from the origin. The orientation of the particle is given by the angle $\phi$ of the $l=0$ leg to the $x$-axis, so that its unit vector is, $\textbf{e}_0 \equiv (\cos \phi, \sin \phi)$. 

Imagine rotating the particle clockwise through a full turn. As the particle is rotated one particle leg will eventually  collide with one of the constraining points. Let us define the vector $\textbf{a}_l$ as the position vector of the constraint which the $l$th leg \textit{first} crosses during this clockwise rotation. In a similar way we define, $\textbf{b}_l$ as the position vector of the corresponding point for the anti-clockwise rotation. The important feature of these vectors is their angle to the $x$-axis defined in terms of the components of the vectors as,
\begin{equation}
\begin{split}
 \tan\alpha_l = \frac{a^{l}_y}{a^{l}_x} \ \ \ \text{and} \ \ \ \tan\beta_l = \frac{b^{l}_y}{b^{l}_x}.
\end{split}
\end{equation}
Further, we define the angles indicated in Fig.(\ref{fig:2DParticles}a), 
\begin{equation}
\label{constraints}
	\eta_l = \phi + \Delta_l - \alpha_l \ \ \ \text{and} \ \ \ \theta_l = \beta_l - \phi - \Delta_l,
\end{equation}
where $\Delta_l$ is the angle between the $l=0$ leg and the $l$th leg. The particle is constrained by the points which are closest to it, so the range of accessible angles is, 
\begin{equation}
\Omega = \min_{l,m} \ \eta_l + \theta_m .
\end{equation}
To obtain the free energy as given in equation (\ref{Fenergy1}), an average needs to be taken over the distribution of constraints, which means averaging over all the angles, $\eta_l$ and $\theta_m$. The probability distribution, $P(\eta_l)$ and $P(\theta_m)$, can be determined from a simple argument. For any given area of size $a$, the average number of points inside is the same, $\nu a$, and independent of the number found in a different area. Therefore, the points are distributed according to a Poisson distribution. When the particle rotates through an angle $\eta_l$, say, the $l$th leg sweeps out an area of size $L^2 \eta_l /2$. The probability that there are no constraining points in this area is, $P_0(\eta_l) = \exp(-\nu L^2 \eta_l/2)$. By definition the $l$th leg \textit{firsts} collide with a constraint after rotating clockwise through an angle $\eta_l$, so it follows that $P(\eta_l) = - P_0'(\eta_l)$. So we find, 
\begin{equation}
P(\eta_l) = \frac{\nu L^2}{2} e^{-\nu L^2 \eta_l /2},
\end{equation}
and exactly the same form for $P(\theta_m)$.

When the shear strain is applied, the constraining points all move and the angles $\alpha_l$, $\beta_l$, $\eta_l$ and $\theta_l$ all change. We assume the the constraining points are displaced affinely when the shear is applied, therefore the position vectors, $\textbf{a}_l$ and $\textbf{b}_l$, transform according to the rule,
\begin{equation}
\label{vectransform}
\textbf{a}_l \to \tilde{\textbf{a}_l} = (\mathbb{I} + \kappa) \cdot \textbf{a}_l,
\end{equation}
with the same expression for $\tilde{\textbf{b}}_l$. For the simple $x$-$y$ shear we consider, the new angles can be determined from, 
\begin{equation}
\label{transalpha}
\tan \tilde{\alpha}_l = \frac{a^{l}_y}{a^{l}_x+ \gamma a^{l}_y} = \frac{\tan \alpha_l }{1 + \gamma \tan \alpha_l},
\end{equation}
and an analogous expression for $\tilde{\beta}_l$. We focus on small strains with $\gamma \ll 1$. Expanding (\ref{transalpha}) in this limit, we find, 
\begin{equation}
\label{transalpha2}
\tilde{\alpha}_l = \alpha_l - \gamma \sin^2 \alpha_l + \gamma^2 \sin^3 \alpha_l \cos \alpha_l + \mathcal{O}(\gamma^3),
\end{equation} 
and similarly for $\beta_l$. The terms to second order in $\gamma$ will contribute to the linear elasticity. For the non-linear response, more terms must be retained. The angles, $\tilde{\eta}_l$ and $\tilde{\theta}_l$ are then found from, 
\begin{equation}
\label{transetatheta}
\tilde{\eta}_l = \phi + \Delta_l - \tilde{\alpha}_l \ \ \ \text{and} \ \ \ \tilde{\theta}_l = \tilde{\beta}_l - \phi - \Delta_l,
\end{equation}
which leads to the transformed range of accessible angles, 
\begin{equation}
\label{transomega}
\tilde{\Omega} = \min_{l,m} \ \tilde{\eta}_l + \tilde{\theta}_m.
\end{equation}
This model forms the basis of a simple numerical simulation we use for comparison to our analytic results. This computes the accessible angle and the free energy numerically by the following simple procedure. First, a particle is placed in a random orientation in the plane. Then the obstacles are placed at random on a circle surrounding the particle; the number of obstacles is proportional to the density, $\pi \nu L^2$. The two closest obstacles to the particle are then found and their positions determine the accessible angle. The positions of the obstacles on the circle are then all moved according to (\ref{transalpha}). The logarithm of the ratio of the accessible angle before and after the transformation gives the free energy for that configuration. These steps are repeated for many different configurations of the particle and obstacles and the average free energy over these realisations is calculated. This essentially calculates (\ref{Fenergy1}) numerically using a Monte-Carlo integration for the averages. 
\subsection{Rods}
The first shape we consider is a rod of length $2L$, which we can think of as two legs of length $L$ which are anti-parallel. The unit vectors giving the orientations of these legs are therefore, $\textbf{e}_0 = (\cos \phi, \sin \phi) = - \textbf{e}_1$. Due to the symmetry of the rod and the shear flow considered, it does not matter which leg of the rod is constrained. This means we only need to consider the angles $\eta_0$ and $\theta_0$, from which we will drop the subscripts in this section. Using equations (\ref{transalpha2}) and (\ref{transetatheta}), we can determine the transformed angles $\tilde{\eta}$ and $\tilde{\theta}$, 
\begin{equation}
	\tilde{\eta} = \eta + \gamma \sin^2(\phi - \eta) - \gamma^2 \sin^3 (\phi - \eta) \cos (\phi - \eta),
\end{equation}
\begin{equation}
	\tilde{\theta} = \theta - \gamma \sin^2(\phi + \theta) + \gamma^2 \sin^3 (\phi + \theta) \cos (\phi + \theta).
\end{equation}
The transformed range of accessible angles is then calculated from $\tilde{\Omega} = \tilde{\eta}+\tilde{\theta}$, 
\begin{equation}
\begin{split}
	&\tilde{\Omega} =\Omega + \gamma\sin(\eta + \theta) \sin(\eta - \theta - 2\phi )-\frac{\gamma^2}{8} \bigg[ \sin\big(4(\eta - \phi)\big)\\
	&+ \sin\big(4(\theta + \phi)\big) -4 \cos(\eta - \theta - 2\phi) \sin(\eta + \theta)  \bigg].
\end{split}
\end{equation}
For convenience let us define the functions, $f(\eta,\theta ; \phi)$ and $g(\eta,\theta ; \phi)$ such that,
\begin{equation}
\tilde{\Omega} = \Omega(1 + \gamma f + \gamma^2 g /2 ).    
\end{equation}
To second order in $\gamma$, the entropy of the test particle for this realisation is,
\begin{equation}
S = S_0 + k_B \gamma f + \frac{k_B \gamma^2}{2}(g - f^2),
\end{equation}
where $S_0 = k_B \log \Omega$ is the entropy of the undeformed state. Now the free energy to the same order is according to (\ref{Fenergy1}), 
\begin{equation}
\begin{split}
	F = F_0 - \frac{k_B T}{2\pi}\rho\int_{0}^{2\pi} d\phi \ \bigg\langle\gamma f + \frac{\gamma^2}{2}(g - f^2)\bigg\rangle,
\end{split}
\end{equation}
where $F_0$ is the free energy of the un-deformed state which is an irrelevant constant we may drop. The distribution function for $\phi$ is taken to be uniform and the average of the constraints is defined as,
\begin{equation}
	\big\langle \cdots \big\rangle = \int_{0}^{\pi} d\theta P(\theta) \int_{0}^{\pi} d\eta P(\eta) \big(\cdots\big).
\end{equation}
It is straightforward to show that the integrals over $\phi$ of $f$ and $g$ vanish, leaving,
\begin{equation}
\begin{split}
F &=  \frac{k_B T}{8}\rho \gamma^2 (\nu L^2)^2 \int_{0}^{\pi} d\theta d\eta \ e^{-\nu L^2 (\theta +\eta)/2} \frac{\sin^2(\eta + \theta)}{(\eta + \theta)^2} \\
&\times \frac{1}{2\pi}\int_{0}^{2\pi}d\phi \ \sin^2(\eta - \theta - 2\phi ).
\end{split}
\end{equation}
The $\phi$ average is simply $1/2$. To compute the integrals over $\eta$ and $\theta$, we make the substitutions $x = \nu L^2 \eta$ and $y = \nu L^2 \theta$. We can then use the fact that $\nu L^2 \gg 1$ to replace the upper integration limits by $\infty$. 
\begin{equation}
	F = \frac{k_B T}{16} \rho\gamma^2 \int_{0}^{\infty} dx dy \ e^{-(x+y)/2} \frac{\sin^2\big(\frac{x}{\nu L^2} + \frac{y}{\nu L^2} \big)}{\big(\frac{x}{\nu L^2} + \frac{y}{\nu L^2}\big)^2}.
\end{equation}
This can be calculated for $\nu L^2 \gg 1 $, see appendix \ref{app:rodint}. The final answer is
\begin{equation}
\label{finalFErod}
F \approx \frac{k_B T\gamma^2}{4}\rho\bigg(1 - \frac{32}{3(\nu L^3)^2}\bigg).
\end{equation}
This directly leads to the linear elastic stress using (\ref{virtualwork2}), 
\begin{equation}
\label{finalstressrod}
	\sigma_0 (\nu) \sim \frac{k_B T}{2}\rho \gamma \bigg(1 - \frac{32}{3(\nu L^2)^2}\bigg),
\end{equation}
The important features of this expression are that it is approximately independent of density and it agrees with a calculation based on a reptation theory. 

In the case of rod shaped particles, it is possible to compute the elastic stress by an alternative argument. The main details of this textbook derivation are left to appendix \ref{app:reptrodcalc}; here we summarise the main results. The relevant component of the elastic stress is found to be, 
\begin{equation}
\label{elasticresponserods}
\sigma_{xy} = \frac{k_B T}{2} \rho \gamma \begin{cases}   
 e^{-4 D_r^0 t} \ , & \text{for} \ t < \tau_{\text{int}} \\
 e^{4 \tau_{\text{int}} (D_r(\nu) - D_r^0)} e^{-4 D_r(\nu) t} \ , & \text{for} \ t > \tau_{\text{int}}
\end{cases}.
\end{equation}
where $D_r^0$ and $D_r(\nu)$ are the rod's rotational diffusion coefficients in the absence and presence of constraints respectively. The latter is estimated using the reptation argument such that, $D_r(\nu) \sim D_r^0 (\nu L^2)^{-2}$. As the average angle between constraints is $\sim (\nu L^2)^{-1}$, the time taken for the rod to interact with the constraints is approximated by, $D_r^0 \tau_{\text{int}} \sim (\nu L^2)^{-2}$. This leads to the initial elastic stress,
\begin{equation}
	\sigma_0(\nu) \sim \frac{1}{2} k_B T \rho\gamma e^{- b/(\nu L^2)^{2}} \sim \frac{1}{2} k_B T \rho\gamma \bigg(1 - \frac{b}{(\nu L^2)^2} \bigg),
\end{equation}
where $b$ is a positive constant. This is precisely the same form as in (\ref{finalstressrod}), found by our geometric argument. The reptation argument does not predict the value of $b$ exactly and, due to the approximations used, we do not expect the value of $32/3$ in (\ref{finalstressrod}) to be precise either. However, the prefactor of $1/2$ and the scaling with density found by our new geometric method agree precisely with those found using the well known techniques. 
\subsection{Wire Frame Particles}
\label{sec:BentParts}
We now consider the simplest variant to a rod, a bent particle with two legs joined at an angle $\chi$. In the framework outlined previously we have $\Delta_l = l \chi$ in (\ref{constraints}) and $l = 0,1$. 
The free energy is then calculated by essentially the same procedure as for rods, the only difference being that we need to deal with different possible combinations of constraints. This is done by weighting the different contributions by their appropriate probabilities, such that the free energy is, 
\begin{equation}
\begin{split}
	F =& - \frac{k_B T}{2\pi}\rho \int_{0}^{2\pi}d\phi \sum_{l,m=0,1} \int_{0}^{\pi}d\eta_{l} P(\eta_l) \int_{0}^{\pi} d\theta_{m} P(\theta_{m})\\
&\int_{\eta_l}^{\pi} d\eta P(\eta) \int_{\theta_{m}}^{\pi} P(\theta)\log(\tilde{\eta}_l + \tilde{\theta}_{m}).  
\end{split}
\end{equation}
Making use of the same substitutions as before, $x_j = \nu L^2 \eta_j$ and $y_j = \nu L^2 \theta_j$, the integrals on $\eta$ and $\theta$ can be evaluated simply, with each giving a factor of $e^{-\eta_l/2}$ and $e^{-\theta_m/2}$ respectively. The expansion in (\ref{transalpha2}), along with the definitions (\ref{transetatheta}), are then used to expand the logarithm to second order in $\gamma$, from which it is straightforward to show that the only contribution which will not vanish when integrated over $\phi$ is,
\begin{equation}
\label{FErefpoint}
\begin{split}
&\frac{k_B T \gamma^2}{16\pi}\rho\int_{0}^{2\pi}d\phi \sum_{l,m=0,1} \int_{0}^{\infty}dx_{l} \int_{0}^{\infty} dy_{m} \frac{(\nu L^2)^2 e^{-(x_l+y_m)}}{(x_l + y_m)^2}\\
&\times \bigg[\sin^2\Big(\phi + l \chi- \frac{x_l}{\nu L^2}\Big) - \sin^2\Big(\phi + m \chi + \frac{y_m}{\nu L^2}\Big)\bigg]^2.  
\end{split}
\end{equation}
At this stage we can drop the subscripts on $x$ and $y$, and take the sums and $\phi$ integral to give, 
\begin{equation}
\label{FEnLDiv}
\begin{split}
F=\frac{k_B T \gamma^2}{16}\rho\int_{0}^{\infty}&dx dy \frac{(\nu L^2)^2 e^{-(x+y)}}{(x + y)^2}\\
&\times \bigg[1- \cos^2 \chi \cos\bigg(\frac{2(x+y)}{\nu L^2}\bigg)\bigg].
\end{split}
\end{equation}
At first, this integral appears to diverge because of the inclusion of the point $x=y=0$. However, this point can be omitted for physical reasons. In any real system there will be a finite excluded volume. This leads to the integral being cut off at small values, making it finite. The qualitative behaviour of the integral can be understood by `pre-averaging' the constraints so that the angles, $\eta$ and $\theta$, are replaced by their average values, $\langle \eta \rangle = \langle \theta \rangle = 2(\nu L^2)^{-1}$. This sets $x \sim y \sim 1$ in (\ref{FEnLDiv}). Expanding the resulting expression in powers of $\nu L^2$ gives the free energy as, 
\begin{equation}
\label{FEnLFinal}
\begin{split}
&F = \frac{k_B T \gamma^2}{2} \rho \Big[a \ (\nu L^2)^2 (1 - \cos^2 \chi) + b \ \cos^2 \chi\Big],
\end{split}\
\end{equation} 
where $a$ and $b$ are positive constants. In appendix \ref{app:rigint}, we treat the excluded volume cut off more rigerously, and find that the scaling found in (\ref{FEnLFinal}) has only logarithmic corrections which depend on the exact cutoff. Therefore, without losing qualitative accuracy, we take the coefficients $a$ and $b$ to be unknown fitting parameters. 

The stress follows easily as,  
\begin{equation}
	\sigma_0(\nu) = \rho \Big[a \ (\nu L^2)^2 (1 - \cos^2 \chi) + b \ \cos^2 \chi\Big] k_B T \gamma.
\end{equation}
This can show a concentration dependence very different from that of rods, depending on the size of $\chi$. For this different behaviour to be seen we require, 
\begin{equation}
	(\nu L^2)^2 (1 - \cos^2 \chi) \gtrsim \cos^2 \chi.
\end{equation}
After some simple re-arrangement this is, 
\begin{equation}
\tan \chi \gtrsim \frac{1}{\nu L^2},
\end{equation}
which, since $\nu L^2 \gg 1$, implies that if the particle is bent through an angle much larger than $\chi_c \sim (\nu L^2)^{-1}$, the dominant scaling of the stress will be $\propto (\nu L^2)^2$. This means that even modestly bent particles ($\chi \sim \mathcal{O}(1)$) will have a significantly different elastic response than rods. 

In particular if we consider `L-shaped' particles, where $\chi = \pi/2$ the elastic stress is simply, 
\begin{equation}
\label{finalstressL}
	\sigma_0(\nu) \propto k_B T \rho (\nu L^2)^2 \gamma.
\end{equation}
This is also observed for other symmetric, N-legged shapes with the legs all joined at the same angle, $2\pi/N$. Examples of these particles are Y-shapes and crosses have been fabricated in practice from double stranded DNA \cite{Xing2018MicrorheologyHydrogels, Biffi2013PhaseNanostars}. It is interesting to note that these special, symmetric shapes have \textit{no} elasticity in dilute suspension for fundamental symmetry reasons\cite{PhysRevE.102.032615}. Suspensions of these particle shapes therefore will be most sensitive to concentration changes. This has important implications for designing these DNA systems as functional nano-materials, where it is desirable to have highly tuneable, concentration or connectivity dependent properties. 

\begin{figure}\includegraphics[width=8cm]{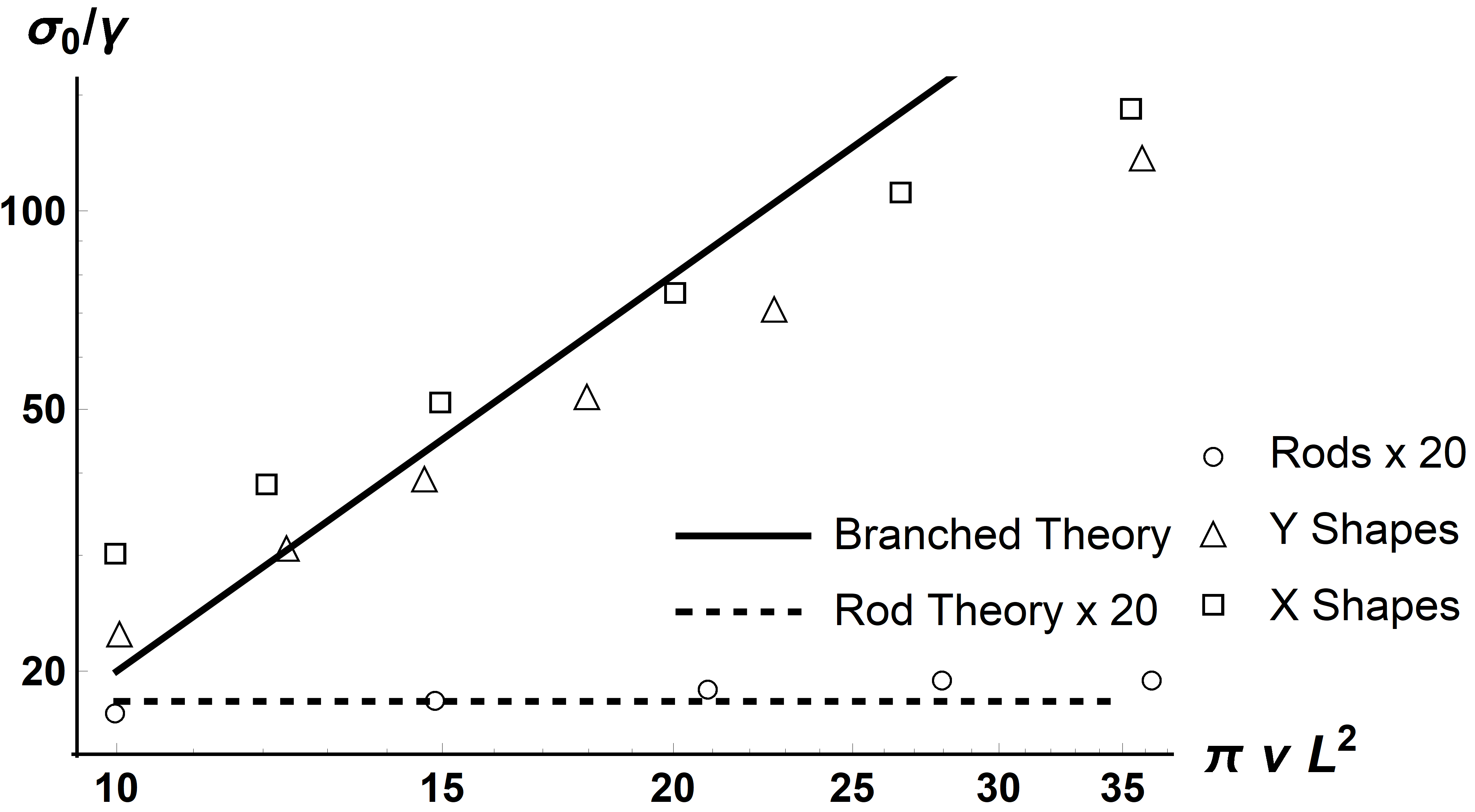}
	\caption{\label{fig:StressScaling} A log-log plot of the elastic stress per particle $\sigma_0$ as a function of density, $\nu$, for rods, Y and X shaped particles. The points represent the simulation data; circles for rods, triangles for Y-shapes and squares for X-shapes. Note that the data for rods has been multiplied by 20. The dashed line is the theoretical result for rods (multiplied by twenty) and the solid line the scaling theory for branched or bent particles, with the intercept fitted. There is clear qualitative agreement between the theory and the data. For rods the stress is small and approximately independent of density, whereas for branched particles the stress is much larger and increases rapidly with the density. On the other hand there is a quantitative discrepancy. The predicted scaling is $\propto \nu^2$, but the fitted exponent for the data is smaller $\sim 1.4$. This can be explained by deviations from the linear elasticity at very small strains in these systems.}
\end{figure}

Equations (\ref{finalstressL}) and (\ref{finalstressrod}) show that wire frames behave in a strikingly different way compared to rods. As $\nu L^2 \gg 1$, we predict that the elastic stress for wire frames is significantly larger than that of rods, and very sensitive to the concentration of constraining points. These results are qualitatively consistent with the simulations, see Fig(\ref{fig:StressScaling}). Here the simulation data for the stress per particle is plotted as a function of density for rods, Y-shapes (N=3 stars) and X-shapes (N=4 stars) and compared to the theoretical predictions (note that the data for rods has been multiplied by 20 for ease of visualisation). The stress is calculated in the simulations by fitting the free energy to a quadratic for strains between $\gamma = 0.02$ and $0.1$.

There is, however, a quantitative difference between the predicted scaling and that of the data. The stress for branched particles is predicted to scale as $\nu^2$, but the fitted exponent for the data is noticeably smaller, $\sim 1.4$. In the simulations particles which bend are removed from the system, i.e the bending modulus $K=0$. As will be shown later (section \ref{sec:nonlinear}), this leads to strong shear thinning in these systems even for relatively small strains, leading to the apparently smaller stress than predicted by the linear theory. 
\section{General Method: Non-Linear Response}
\label{sec:nonlinear}
\begin{figure*}\includegraphics[width=17.2cm]{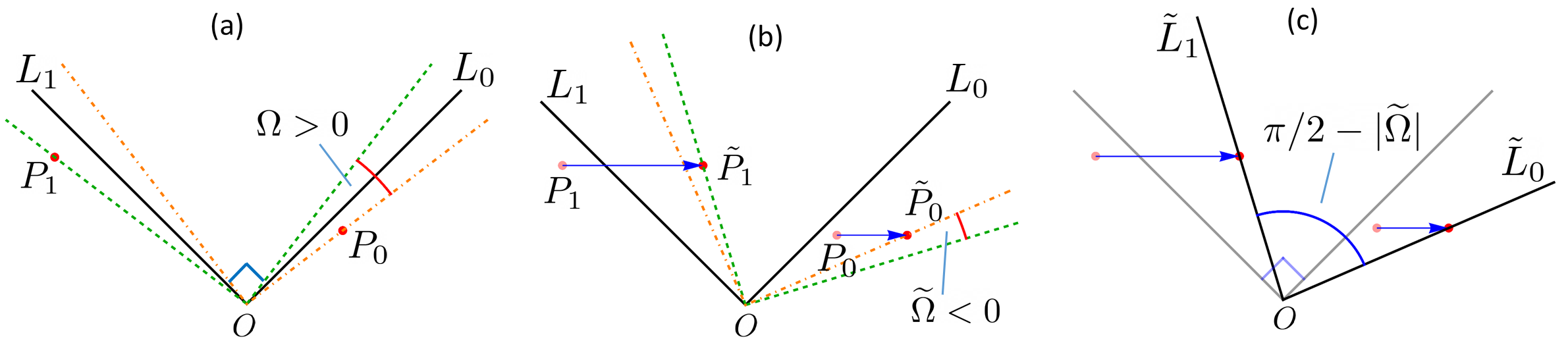}
	\caption{\label{fig:BendingMech} A sketch of the bending mechanism for an L-Shaped wire frame. (a) An undeformed configuration. The clockwise rotation of the L-shaped particle is constrained by the point, $P_0$, which hinders the leg, $OP_0$. The anti-clockwise rotation is constrained by the point $P_1$ hindering leg $OL_1$. The L-shape can only rotate between the orange line, $OP_0$, and the green line, $OP_1$. The accessible configuration space is the angle between these two lines, $\Omega$. (b) When the step strain is applied, the points $P_0$ and $P_1$ are displaced  to $\tilde{P}_0$ and $\tilde{P}_1$ respectively. The L-shape must now lie between the lines $O\tilde{P}_0$ and $O\tilde{P}_1$. The line $O\tilde{P}_1$ is now to the right of $O\tilde{P}_0$, therefore the L-shape cannot keep its original shape and $\tilde{\Omega} < 0$. (c) To satisfy the constraints, the particle must deform from its original state. Assuming its two legs stay straight, the angle between them becomes $\pi/2 - |\tilde{\Omega}|$.}
\end{figure*}

We now extend our approach to study the non-linear elasticity of these systems. To calculate this we seek an expression for the free energy accurate to order $\gamma^4$. In addition to expanding the transformed entropy, $\log\tilde{\Omega}$, to this order, we need to introduce another physical process; the deformation of the particles themselves. For example, as sketched in Fig.(\ref{fig:BendingMech}), an L shaped particle is forced to bend when the surrounding constraining points are displaced by the shear deformation. This situation cannot occur for straight rods, but can for general wire frame particles.

To demonstrate this, consider two lines at angles $\alpha$ and $\beta$ to the $x$-axis with $\alpha < \beta$. These represent the constraints placed on a particular particle; the particle is trapped between these two lines. The particle will need to bend if, after the deformation, these lines cross. For this to happen we must be able to find a pair of angles, $\alpha \neq \beta$, for which $\tilde{\alpha} = \tilde{\beta}$. Initially, take the particle to be a rod. This means that both lines will transform in the same way, according to (\ref{transalpha}), which can be re-arranged to, 
\begin{equation}
\label{cotalpha}
	\cot \tilde{\alpha} = \cot \alpha + \gamma.
\end{equation} 
Subtracting this from the equivalent expression for $\tilde{\beta}$, and insisting that the transformed angles need to be the same we find that $\cot \alpha = \cot \beta$, which implies, $\alpha = \beta + n \pi$. Due to the symmetry of the particle shape, rotation of one of the constraining angles by an integer multiple of $\pi$ results in the same constraint. Therefore, for rods, $\tilde{\alpha} = \tilde{\beta}$ if and only if $\alpha = \beta$ and hence, bending is impossible. 

For L-shaped particles, if the constraints are placed on different legs, the transformation rule for one of the angles is different. Taking $\beta$ to be a constraint on one leg and $\alpha$ to be on the other, then  $\beta$ will transform as if it were rotated by $\pi/2$ with respect to $\alpha$. Hence $\tilde{\beta}$ is given by, 
\begin{equation}
	\cot \tilde{\beta} = \gamma - \tan \beta. 
\end{equation}
Again, we subtract this from (\ref{cotalpha}) and insist $\tilde{\alpha} = \tilde{\beta}$ which yields, $\cot \alpha = \tan \beta$. This has solutions, $\beta = \alpha - \pi/2 + n \pi$. Clearly it is possible to find valid solutions with $\alpha \neq \beta$ and so for these particles, bending is possible for certain configurations. 

To deal with the possibility of bending we introduce the function, $P(\gamma;\Gamma)$. This is the proportion of particles with orientation $\Gamma$ \textit{not} bent at strain $\gamma$. The general form of this function can be determined from simple considerations. When the system is not strained, none of the particles are bent, hence $P(0 ; \Gamma) = 1$. The behaviour of the system is symmetric under a change in sign of $\gamma$, which implies, $P(\gamma;\Gamma) = P(-\gamma;\Gamma)$. The size of strain is taken to be small so that this function is expanded to give, 
\begin{equation}
\label{bendingprob}
	P(\gamma ; \Gamma) = 1 - \gamma^2/ \gamma_c^2(\Gamma),
\end{equation} 
where we have defined the critical strain $\gamma_c(\Gamma)$, whose scaling with density will be determined in the following section. This critical strain should be interpreted as the strain above which all particles in orientation $\Gamma$ are bent. This strain also indicates roughly the limit of accuracy of the treatment presented here, as above this strain additional processes not taken into account will need to be considered, such as the non linear elasticity or potential breakage of the particles themselves.

We may also consider the orientational average of (\ref{bendingprob}), 
\begin{equation}
	\label{bendingprobavg}
	P(\gamma) = 1 - \gamma^2/ \gamma_c^2.
\end{equation} 
This should be interpreted as the total fraction of particles which have not bent at strain $\gamma$. We will use this for comparison to simulations through the scaling of $\gamma_c$ with the density. 

When a particle bends, it stores some elastic energy. This must contribute to the free energy of the system and $P(\gamma ; \Gamma)$ can be used to determine this bending contribution. The probability that a particle \textit{first} bends at a strain $\gamma$ is $- \partial P / \partial \gamma$. If the applied strain is $\gamma$, but a particle in orientation $\Gamma$ first bent at $ \gamma' < \gamma$, then the particle must bend by an angle $\vartheta = (\gamma - \gamma') \Delta(\Gamma)$, where $\Delta$ depends on the particle shape. If the bending modulus of the particle is $K$, the energy associated with this bending is, $K \vartheta^2/2$. The bending contribution to the free energy from one particle in orientation $\Gamma$ is given by; 
\begin{equation}
\label{bendfreeenergy}
F_B(\gamma ; \Gamma)=\frac{K}{2} \rho \Delta^2(\Gamma) \int d\gamma' \ (\gamma - \gamma')^2 \bigg(-\frac{\partial P(\gamma ; \Gamma)}{\partial \gamma}\bigg\lvert_{\gamma'}\bigg),
\end{equation}
where all the possible strains at which the particle could first bend have been summed over, weighted by the appropriate probability. We can now write an expression for the free energy per unit volume which includes both the entropic contribution and the bending contribution.  
\begin{equation}
\label{FreeEnergy}
\begin{split}
&F(\gamma) =  -k_{B} T \rho \int d \Gamma \ \psi(\Gamma) P(\gamma ; \Gamma) \big\langle\log \Tilde{\Omega}(\Gamma ; \gamma)\big\rangle \\
&+ \frac{K}{2} \rho \int d\Gamma \ \psi(\Gamma) \Delta^2(\Gamma) \int_{0}^{\gamma} d\gamma' \ (\gamma - \gamma')^2 \bigg(-\frac{\partial P}{\partial \gamma}\bigg\lvert_{\gamma'}\bigg).   
\end{split}
\end{equation}
The orientation needs to be averaged over the distribution function $\psi(\Gamma)$, which for our case can be taken to be uniform. The first term in the above equation represents the entropic contribution, $F_S(\gamma)$. The factor of $P$ needs to be included since a particle which bends is forced to be in one configuration by the constraints, and as such does not contribute entropically.

The form of the free energy (\ref{FreeEnergy}) is the basis of our treatment of the non-linear elasticity in these systems. We will use this to determine the free energy to order $\gamma^4$ first for rods, then for L-shaped particles. The latter shape is taken for ease, but we would find the same features for any bent or branched particle. 
\subsection{Rods}
\label{sec:nonlinearrods}
The extension to non-linear elasticity for rods is straightforward. As discussed previously, there is no possibility for bending, so $P(\gamma ; \Gamma) = 1$, and all that needs to be done is to expand $\tilde{\Omega}$ and the entropy to order $\gamma^4$. After lengthy but straightforward algebra, the first and third order contributions to the free energy can be shown to vanish after integrating over $\phi$. The second order contribution is that found previously in (\ref{finalFErod}). The fourth order contribution is, after averaging on $\phi$,
\begin{equation}
	\bigg\langle\frac{\gamma^4 \sin^3(\eta + \theta)}{32 (\eta+\theta)^4}\Big[4 (\eta+\theta)\cos(\eta + \theta) + \big(2(\eta+\theta)^2 - 3\big) \Big]\bigg\rangle.
\end{equation} 
The averages over $\eta$ and $\theta$ are then taken in the same way as before, and the free energy to 4th order in $\gamma$ is obtained, 
\begin{equation}
	\label{nonlinFErod}
	\frac{F(\gamma)}{\rho k_B T} = \frac{\gamma^2}{4}\bigg(1 - \frac{c}{(\nu L^2)^2}\bigg) -  \frac{\gamma^4}{32}\bigg(1 - \frac{d}{(\nu L^2)^4}\bigg),
\end{equation}  
where $c$ and $d$ are positive constants. The elastic stress is, dropping the subdominant terms for $\nu L^2 \gg 1$, 
\begin{equation}
	\label{nonlinearstressrod}
\sigma_0 =k_B T \rho \bigg( \frac{1}{2} \gamma - \frac{1}{8} \gamma^3 \bigg).
\end{equation}
This shows that the suspension of rods shear thins, as the co-efficient of the cubic term is negative. This is consistent with the results of a more detailed theory for this system based on the reptation model \cite{Doi1986TheDynamics}, and serves as a useful reference for comparison to the behaviour of the L-shapes.  
\subsection{L-Shapes}
\label{sec:bendingLs}
For L-shapes, bending is possible since configurations can be found where $\tilde{\Omega} \leq 0$. To compute $P(\gamma ; \phi)$ for these shapes, we first find the probability that a particle with orientation $\phi$ \textit{has} bent, 
\begin{equation}
	P_B(\gamma ; \phi ) = 1 - P(\gamma ; \phi).
\end{equation}
This is the probability that $\tilde{\Omega} \leq 0$ which, from the previous discussion, happens only when the constraints are placed on different legs. Therefore there are two possibilities for bending to occur, 
\begin{equation}\label{regions}
	\begin{split}
	\text{(I)} \ \ &\tilde{\eta}_1 < \tilde{\eta}_0 \ \ , \ \ \tilde{\theta}_0 < \tilde{\theta}_1 \ \ \text{and} \ \ \tilde{\eta}_1 + \tilde{\theta}_0 \leq 0,\\
	\text{(II)} \ \ &\tilde{\eta}_0 < \tilde{\eta}_1 \ \ , \ \ \tilde{\theta}_1 < \tilde{\theta}_0 \ \ \text{and} \ \ \tilde{\eta}_0 + \tilde{\theta}_1 \leq 0.
	\end{split}
\end{equation}
The probability of bending is then the sum of the integrals of the constraint probabilities over each of these regions. We give the calculation for region (I) in detail, the steps are the same for (II) so we will simply state the result. The integral over region (I) is formally, 
\begin{equation}
\label{PB1}
\begin{split}
	P_B(\gamma ; \phi) = &\int_{\tilde{\eta}_1 + \tilde{\theta}_0 \leq 0} d\eta_1 d\theta_0 \ P(\eta_1) P(\theta_0) \\
	 &\times \int_{\tilde{\eta}_0 > \tilde{\eta}_1} d\eta_0 \ P(\eta_0) \int_{\tilde{\theta}_1 > \tilde{\theta}_0} d\theta_1 \ P(\theta_1). 
\end{split}
\end{equation}
To make progress we require expressions for the transformed constraining angles, which are deduced straightforwardly from previous definitions,
\begin{equation}
\label{transanglshapes}
\begin{split}
&\tilde{\eta}_0 = \eta_0 + \gamma \sin^2(\phi - \eta_0) - \gamma^2 \sin^3(\phi - \eta_0) \cos(\phi - \eta_0),\\
&\tilde{\eta}_1 = \eta_1 + \gamma \cos^2(\phi - \eta_1) + \gamma^2 \cos^3(\phi - \eta_1) \sin(\phi - \eta_1),\\
&\tilde{\theta}_0 = \theta_0 - \gamma \sin^2(\phi + \theta_0) + \gamma^2 \sin^3(\phi + \theta_0) \cos(\phi + \theta_0),\\
&\tilde{\theta}_1= \theta_1 - \gamma \cos^2(\phi +\theta_1) - \gamma^2 \cos^3(\phi + \theta_1) \sin(\phi+\theta_1).
\end{split}
\end{equation}
Let us define the function $A(\theta_0)$ as the value of $\theta_1$ where $\tilde{\theta}_1 = \tilde{\theta}_0$. The function $\tilde{\theta}_1$ in (\ref{transanglshapes}) is monotonic in $\theta_1$ for small values of $\gamma$, so it follows that if $\theta_1 > A(\theta_0)$, then $\tilde{\theta}_1 > \tilde{\theta}_0$. We also define $B(\eta_1)$ as the value of $\eta_0$ where $\tilde{\eta}_1 = \tilde{\eta}_0$. These functions are found to be, 
\begin{subequations}
\begin{equation}
\begin{split}
A(\theta_0) &= \theta_0 + \gamma \cos(2\theta_0 + 2\phi) \\
&+ \frac{1}{2}\gamma^2\big(\sin(2\theta_0 + 2\phi) - \sin(4\theta_0 + 4\phi)\big),
\end{split}
\end{equation}
\begin{equation}
\begin{split}
B(\eta_1) &= \eta_1 + \gamma \cos(2\eta_1 - 2\phi) \\
&- \frac{1}{2}\gamma^2\big(\sin(2\eta_1 - 2\phi) + \sin(4\eta_1 - 4\phi)\big).
\end{split}
\end{equation}
\end{subequations}
The integrals over $\theta_1$ and $\eta_0$ in (\ref{PB1}) are now, 
\begin{equation}
\label{PB2}
	\int_{B(\eta_1)}^{\pi} d\eta_0 \ P(\eta_0)\int_{A(\theta_0)}^{\pi} d\theta_1 \ P(\theta_1).
\end{equation}
However it can be shown that $A(0) < 0$ and $B(0) < 0$ when, 
\begin{equation}
\label{allowedphi}
	\frac{\pi+\gamma}{4} < \phi < \frac{3\pi + \gamma}{4} \ \ \text{and} \ \ \frac{5 \pi + \gamma}{4} < \phi < \frac{7 \pi + \gamma}{4}. 
\end{equation}
In which case, the lower limits on the integrals in (\ref{PB2}) are both replaced by zero. 

\begin{figure}
	\includegraphics[width=8.6cm]{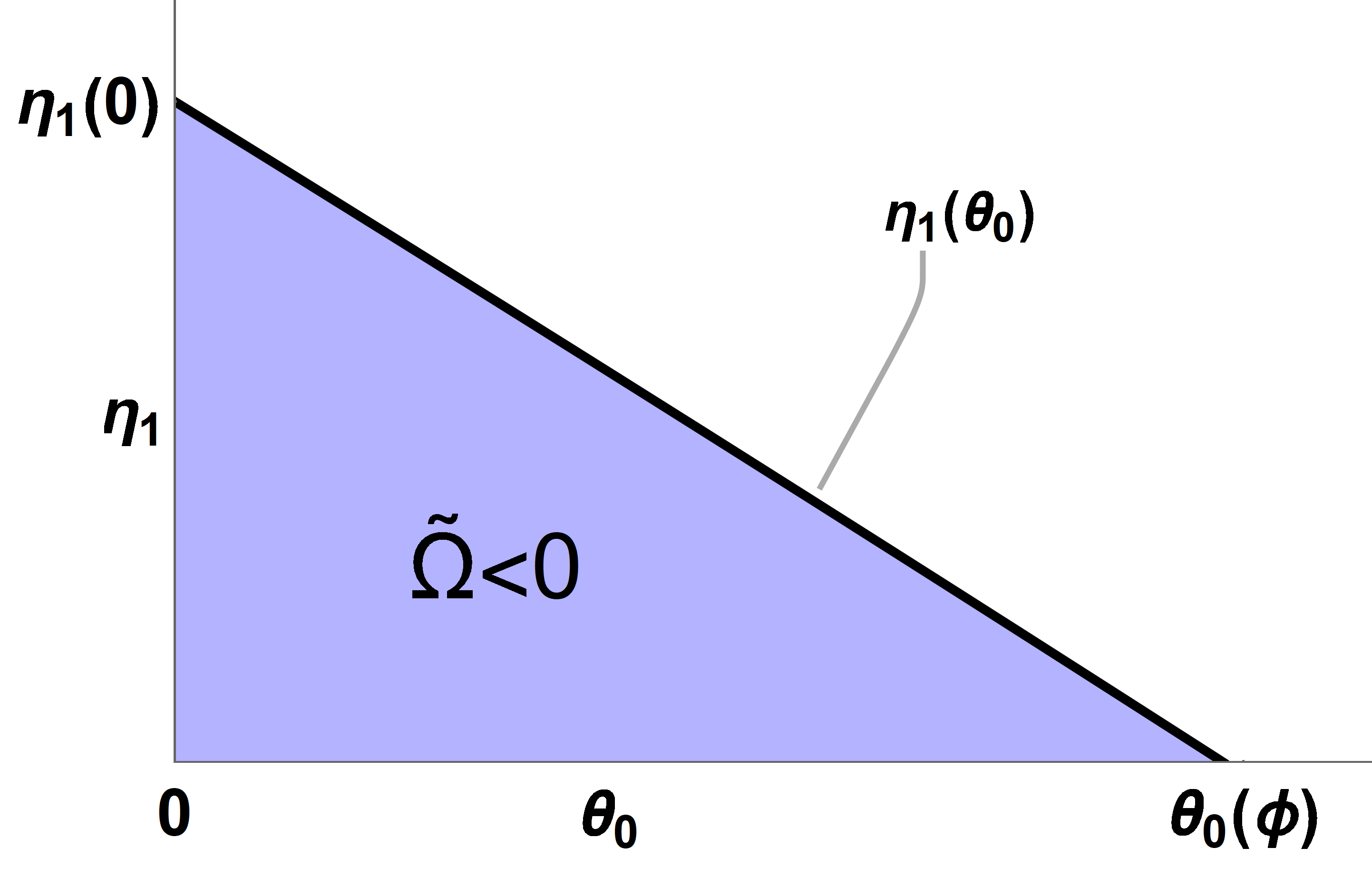}
	\caption{\label{fig:OmegaRegionPlot} A sketch of the $\theta_0,\eta_1$ plane. The line $\eta_1(\theta_0)$ with $\gamma = 0.1$ and $\phi = \pi/2$, is shown. From (\ref{eta1oftheta0}) this must be approximately a line with gradient $-1$, which intersects the two axes at $(0,\theta_0(\phi))$ and $(\eta_1(0),0)$. The small variations about this line are too small to be noticeable on this figure. The shaded region is where $\tilde{\Omega} < 0$.}
\end{figure}

\begin{figure}
	\includegraphics[width=8.6cm]{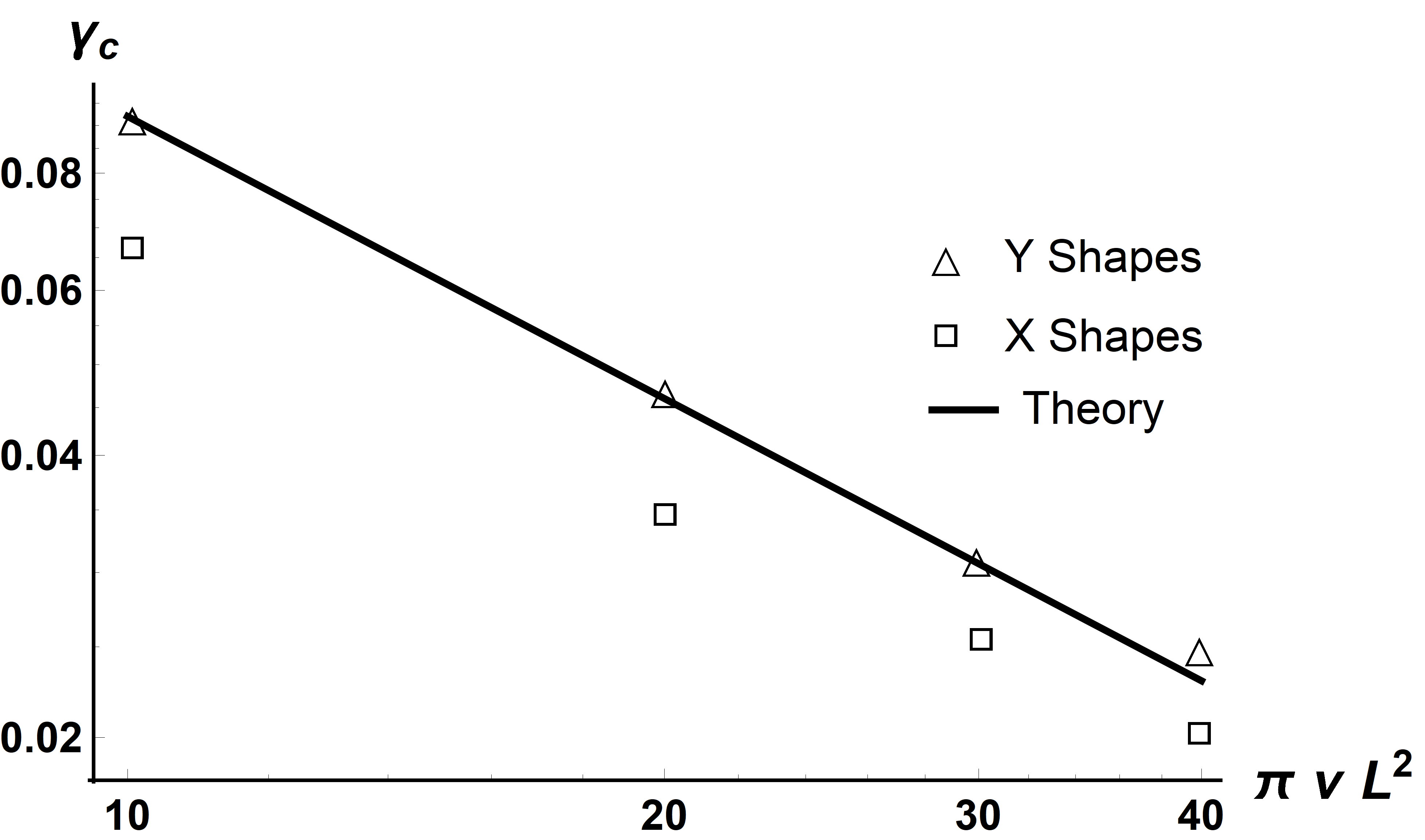}
	\caption{\label{fig:GammaC} A log-log plot of the critical strain, $\gamma_c$, against dimensionless density, $\pi \nu L^2$. The points are data taken from the 2D simulations, triangles for Y-shaped particles and squares for X-shapes. The solid line is the scaling predicted theoretically (with the unknown intercept fitted to the Y-shape data). We see excellent agreement between the predicted scaling and the data.}
\end{figure}

In a similar fashion, we define the function $\eta_{1}(\theta_0)$ where $\tilde{\eta}_1 + \tilde{\theta}_0 = 0$,
\begin{equation}
\label{eta1oftheta0}
\begin{split}
\eta_1(\theta_0) &= - \theta_0 - \gamma \cos(2 \theta_0 + 2\phi) \\
&-\frac{1}{2} \gamma^2 \big(\sin(2\theta_0 + 2\phi) - \sin(4\theta_0 + 4\phi)\big).
\end{split}
\end{equation} 
This defines a line in the $\theta_0$, $\eta_1$, plane, which is approximately a straight line between the points $(0,\eta_1(0))$ and $(\theta_0(\phi),0)$, with gradient negative one, as shown in Fig.(\ref{fig:OmegaRegionPlot}). The shaded region in Fig.(\ref{fig:OmegaRegionPlot}) enclosed by the axes and this line is the region where $\tilde{\Omega} \leq 0$. For certain orientations this region vanishes, and bending is not possible. For the region to exist, it is necessary for $\eta_1 (0) > 0$, which from (\ref{eta1oftheta0}) occurs for orientations satisfying (\ref{allowedphi}). Therefore, the contribution to $P_B$ from region (I) is determined from the integral, 
\begin{equation}
\label{PB3}
\int_{0}^{\theta_0(\phi)} d\theta_0 \int_{0}^{\eta_1(\theta_0)} d\eta_1  \int_{0}^{\pi} d\eta_0d\theta_1 \ P(\eta_0)P(\eta_1)P(\theta_0)P(\theta_1),
\end{equation}
with $\phi$ satisfying (\ref{allowedphi}) and, 
\begin{equation}\label{theta0phi}
	\theta_0(\phi) = -\gamma \cos 2\phi - \frac{\gamma^2}{2} \big(\sin 2\phi + \sin 4\phi\big). 
\end{equation}
Using the substitutions $x_l = \nu L^2 \eta_l$ and $y_l = \nu L^2 \theta_l$ as before, the $\eta_0$ and $\theta_1$ integrals will each evaluate to one and we are left with, 
\begin{equation}
\label{PB4}
\frac{1}{4}\int_{0}^{\nu L^2 \theta_0(\phi)} dy_0 \int_{0}^{\nu L^2 \eta_1(y_0/ \nu L^2)} dx_1 e^{-(x_1 + y_0)/2}.
\end{equation}
This integral is evaluated in appendix \ref{app:bendint}. The result to lowest order in $\gamma$, is,
\begin{equation}\label{bendprobint}
1 - e^{-\nu L^2 \theta_0(\phi)/2} -\frac{1}{2}\nu L^2 \theta_0(\phi) e^{- \nu L^2 \eta_0(0)/2}. 
\end{equation}
Using the definitions of $\theta_0(\phi)$ and $\eta_0(0)$, this is expanded to second order in $\gamma$ to give the contribution to $P_B$ from region (I), 
\begin{equation}
\text{(I)} = \ \ \begin{cases}   
0 , & \text{if} \ \phi \notin \text{(\ref{allowedphi})} \\
\Big(\frac{\nu L^2}{2}\Big)^2 \gamma^2 \cos^2 2 \phi \ , & \text{if} \ \phi \in \text{(\ref{allowedphi})}.
\end{cases}
\end{equation}
The contribution from region (II) is calculated by following the same procedure, and is found to be, 
\begin{equation}
\text{(II)} = \ \ \begin{cases}   
0 , & \text{if} \ \phi \in \text{(\ref{allowedphi})} \\
\Big(\frac{\nu L^2}{2}\Big)^2 \gamma^2 \cos^2 2 \phi \ , & \text{if} \ \phi \notin \text{(\ref{allowedphi})}.
\end{cases}
\end{equation}
The bending probability is simply the sum of these to contributions,
\begin{equation}
	P_{B}(\gamma ; \phi) = \Big(\frac{\nu L^2}{2}\Big)^2 \gamma^2 \cos^2 2 \phi
\end{equation}
Taking the orientational average of this, we obtain the proportion of un-bent particles,
\begin{equation}
	\label{Pbendavg}
	P(\gamma) = 1- (\nu L^2)^2\gamma^2/8.
\end{equation}
From which the scaling of $\gamma_c$ can be extracted, 
\begin{equation}
	\gamma_c \sim \frac{1}{\nu L^2}.
\end{equation}
Even though this was derived explicitly for L-shapes, this scaling is expected for any bent or branched particles. In Fig.(\ref{fig:GammaC}) this scaling is compared to simulation data for Y and X-shaped particles and we see excellent agreement.

The entropic term will need to be expanded to fourth order as it was for rods but with $P(\gamma , \phi)$ included before the orientational average is taken so that the integrals are only taken over the region with $\tilde{\Omega} > 0$. The exact calculation is possible albeit long winded. Fortunately the scaling can be found using the same pre-averaging method as in section \ref{sec:BentParts} (up to logarithmic corrections, see appendix \ref{app:rigint}). We replace the angles $\eta$ and $\theta$ by their average values, and pre-average $\log \tilde{\Omega}$ and $P(\gamma, \phi)$ over $\phi$ in (\ref{FreeEnergy}). After several lines of algebra, we find,
\begin{equation}
	\begin{split}
	&\langle\log \tilde{\Omega}\rangle = -\frac{(\nu L^2)^2}{(16)^2}\gamma^2 -\frac{(\nu L^2)^4}{(16)^4}\gamma^4.
	\end{split}
\end{equation}
The pre-averaged entropic free energy is then found by multiplying this by (\ref{Pbendavg}),
\begin{equation}
	\begin{split}
		&\frac{\bar{F}_S(\gamma)}{\rho k_B T} = \frac{(\nu L^2)^2}{(16)^2}\gamma^2 +\frac{(\nu L^2)^4}{(16)^4}\gamma^4 - \frac{(\nu L^2)^4}{8(16)^2}\gamma^4.
	\end{split}
\end{equation}
Hence, the general form of the free energy to fourth order for bent and branched particles is, 
\begin{equation}
\label{preavgFE}
	\begin{split}
		&\frac{F_S(\gamma)}{\rho k_B T} = \frac{a}{2} (\nu L^2)^2 \gamma^2 - \frac{b}{4} (\nu L^2)^4 \gamma^4,
	\end{split}
\end{equation}
where $a$ and $b$ are positive constants. 

We now compute the bending contribution to the free energy, which requires us to determine the angle through which the particle has bent at a particular strain. At the strain when the particle starts to deform, $\gamma'$, the constraining points are positioned on different legs at $\eta = \theta = 0$. Therefore from (\ref{constraints}), $\alpha = \phi$ and $\beta = \phi + \pi/2$, so that the angle between the two legs is $\chi = \beta - \alpha = \pi/2$. Straining the system further to $\gamma > \gamma'$ changes this angle to $\tilde{\chi} =\tilde{\beta}-\tilde{\alpha}$. This process bends the particle through an angle, $\vartheta = \tilde{\chi} - \pi/2$.  Using previous results, it is straightforward to show that, 
\begin{equation}
	\vartheta = - (\gamma - \gamma') \cos 2 \phi - \frac{1}{2} (\gamma - \gamma')^2 \sin 2 \phi. 
\end{equation}
Only the first term is needed for the bending free energy to fourth order, which is given by
\begin{equation}
	\begin{split}
		F_B(\gamma) &= \frac{1}{4} K (\nu L^2)^2 \rho \int_{0}^{2\pi} \frac{d\phi}{2\pi}\cos^4 2\phi \int_{0}^{\gamma} d\gamma'(\gamma - \gamma')^2 \gamma' \\
		&= \frac{K (\nu L^2)^2}{32} \rho \gamma^4.
	\end{split}
\end{equation}
Finally, we obtain an expression for the free energy for wire frames accurate to order $\gamma^4$, which includes both the entropic and bending contributions,
\begin{equation}
	\frac{F(\gamma)}{\rho k_B T} = \frac{a}{2} (\nu L^2)^2 \gamma^2 + \frac{b}{4} (\nu L^2)^2 \bigg( \frac{K}{k_B T} - c (\nu L^2)^2\bigg) \gamma^4.
\end{equation}
The constants $a,b$ and $c$ are all positive and depend on the particle geometry chosen. The stress immediately follows, 
\begin{equation}
\label{nonlinearstressL}
	\frac{\sigma_0}{\rho k_B T} = a (\nu L^2)^2 \gamma + b(\nu L^2)^2 \bigg( \frac{K}{k_B T} - c (\nu L^2)^2\bigg) \gamma^3.
\end{equation}
This should be compared to the stress for rods (\ref{nonlinearstressrod}). Not only is the stress for the bent particles significantly larger and more sensitive to concentration than for rods, the sign of the cubic term is not necessarily negative. This means that while the rod system will always be shear thinning, the bent particle system can shear harden.  If the elastic modulus of the particles themselves is large enough, 
\begin{equation}
	K \gtrsim (\nu L^2)^2 k_B T,
\end{equation}
the cubic term is positive and the response is shear hardening. Alternatively this condition can be seen as a critical density above which the behaviour transitions from shear hardening to shear softening, 
\begin{equation}
	\nu_c \sim \sqrt{\frac{K}{k_B T L^4}}.
\end{equation}
These conditions are non-trivial, because, while we have assumed $K \gg k_B T$, so the particles are very rigid, we also take $\nu L^2 \gg 1$. If $K/k_B T \sim (\nu L^2)^p$ for any $p \geq 1$, our results are valid but only for $p > 2$ is shear hardening predicted. 

This behaviour has an explanation at the level of the model presented here. When a particle starts to deform, its orientation is completely determined by the surroundings, and as such cannot contribute to the entropic free energy. This effect is captured by the $P(\gamma)$ factor in the first term of equation (\ref{FreeEnergy}). As the applied strain is increased, more and more particles begin to bend, so fewer and fewer contribute entropically. This deficit leads to the shear thinning behaviour of (\ref{preavgFE}). If, at a given strain, the bending contribution is not sufficient to make up this deficit, the total stress will be shear thinning. Therefore, there is some critical value of the bending modulus which must be exceeded to see a shear hardening response. 

Another difference between the behaviour of the wire frames and rods is the relative size of the non-linear term in the stress. For rods both the co-efficients of $\gamma$ and $\gamma^3$ in (\ref{nonlinearstressrod}) are of order one, implying non-linear effects are only important for larger strains. On the other hand, the coefficient of $\gamma^3$ in (\ref{nonlinearstressL}) is $\mathcal{O}((\nu L^2)^4)$ but the coefficient of $\gamma$ is only $\mathcal{O}((\nu L^2)^2)$. This shows that the non-linear effects can begin to be important at very small strains. This can explain the difference between the simulation data and the linear theory in Fig.(\ref{fig:StressScaling}). In the simulations $K=0$, so shear softening is expected, reducing the measured stress compared to what is predicted from the linear theory. 
\section{Discussion}
\label{sec:discussion}
We have discussed the elastic stress response to step strain in a dense suspension of rod-like, bent and branched particles using a simple geometrical method. The principle of this method is to determine the volume of configuration space accessible to a particular particle, given the constraints placed on it by its surroundings, and how this volume changes when the constraints are moved by the applied strain. The accessible volume of configuration space is related to the entropy of the particle, and so when it changes this leads to a change in the free energy of the system. The stress calculated from this free energy should be interpreted as that measured immediately after the step strain is applied.

In this paper we illustrate this method on a simple 2D model system. Here a test particle is surrounded by point-like constraints which it cannot cross. These points play the role of the other particles in three dimensions by providing kinetic constraints. The test particle is free to rotate about its fixed centre until it collides with the constraints. The constraints are taken to transform affinely with the applied strain. We focus specifically on the difference between straight rods and bent or branched wire frames, using L-shaped particles as an example. 

The crucial difference between rods and the other particles is that the constraints on rotation can be placed on different legs. The consequence of this is that the range of angles accessible to an L-shape after the strain is applied can be zero or even negative. This means the particle must have deformed to satisfy the constraints. This must contribute to the free energy. This is taken into account by introducing the function, $P(\gamma)$, interpreted as the proportion of particles which have not bent at a strain $\gamma$. We find that, to lowest order in $\gamma$, this is given by $P = 1 - (\gamma/\gamma_c)^2$, where $\gamma_c$ is the critical strain above which most particles have bent. We determined that $\gamma_c \sim (\nu L^2)^{-1}$, which agrees well with the scaling found from simple simulations of the 2D model. With the inclusion of the bending mechanism, the free energy was found to $\mathcal{O}(\gamma^4)$ and used to discuss the magnitude of the linear elastic stress as well as the nature of the non-linear response for different particle shapes. The linear elastic stress for rods calculated by this method agrees precisely with that found from a calculation based on a reptation model.

This simple model shows two interesting differences between rods and wire frame particles:

1) The linear elastic stress per particle for the wire frames scales proportional to $(\nu L^2)^2$ whereas for rods it is roughly independent of density. This is a significant difference. As we are taking $\nu L^2 \gg 1$, the stress for wire frames is much larger than for rods and much more sensitive to concentration. This stronger scaling is found for any particle bent through an angle $\sim \mathcal{O}(1)$ and so the rheology of the system is highly sensitive to the particle shape. This is in good agreement with simulation data.

2) There is a critical density beyond which the elastic stress for branched particles changes from shear hardening to shear softening. The value of this critical density depends on the elastic modulus of the particle, $\nu_c L^2\sim (K/ k_B T)^{1/2}$. This is also very different from the behaviour for rods, where the system is always shear thinning. 

We hope the simple model presented here may be a useful starting point for investigating other interesting features of wire frame systems, such the decay timescale for the stress in the system of branched particles. For such particles, the reptation process which allows rods to diffuse in a densely constrained system is no longer possible. Hence we may expect to find very different time dependence of the stress and the 2D model provides a framework for investigating this where exact calculations are possible. 
\acknowledgements 
We gratefully acknowledge Prof. Daan Frenkel for many important and insightful discussions. D.A.K. acknowledges financial support from the UK Engineering and Physical Sciences Research Council Ph.D. Studentship award No. 1948692.
\appendix
\section{Evaluation of rod free energy integral}\label{app:rodint}
The free energy for rods to second order in $\gamma$ is given by the integral,
\begin{equation}
	F = \frac{k_B T}{16} \rho \gamma^2 \int_{0}^{\infty} dx dy \ e^{-(x+y)/2} \frac{\sin^2\big(\frac{x}{\nu L^2} + \frac{y}{\nu L^2} \big)}{\big(\frac{x}{\nu L^2} + \frac{y}{\nu L^2}\big)^2}.
\end{equation}
To compute this in the limit $\nu L^2 \to \infty$, we expand the integrand in the variable, $y/(\nu L^2)$, which yields
\begin{equation}
\frac{k_B T}{16} \rho \gamma^2 \int_{0}^{\infty} dx dy \ e^{-(x+y)/2} \bigg[ \frac{\sin^2 (\frac{x}{\nu L^2})}{(\frac{x}{\nu L^2})^2} + C\Big(\frac{x}{\nu L^2}\Big) \frac{y}{\nu L^2}\bigg],
\end{equation}
where $C(a) \equiv 2 \sin(a)(a\cos(a) - \sin(a))/a^3$. Integrating over $y$ gives, 
\begin{equation}
\frac{k_B T}{8} \rho \gamma^2 \int_{0}^{\infty} dx \ e^{-x/2} \bigg[ \frac{\sin^2 (\frac{x}{\nu L^3})}{(\frac{x}{\nu L^3})^2} + \frac{2}{\nu L^2}C\Big(\frac{x}{\nu L^3}\Big)\bigg]. 
\end{equation}
Expanding the integrand in powers of $x/(\nu L^3)$ we finally obtain the free energy as given in equation (\ref{finalFErod}) of the main text,
\begin{equation}
\begin{split}
F &\approx \frac{k_B T}{8} \rho\gamma^2 \int_{0}^{\infty} dx \ e^{-x/2} \bigg(1 - \frac{x(4+x)}{3 (\nu L^2)^2}\bigg) \\
&\approx \frac{k_B T\gamma^2}{4}\rho\bigg(1 - \frac{32}{3(\nu L^3)^2}\bigg).
\end{split}
\end{equation}

\section{Calculation for rods based on reptation}
\label{app:reptrodcalc}
Here we follow a textbook procedure to find the elastic stress for rods. Under $x$-$y$ shear flow, a rod in two dimensions will rotate with angular velocity $\dot{\phi} = - \dot{\gamma}\sin^2 \phi$, about its centre. If the rotational diffusion co-efficient is $\mathcal{D}_r$, then in the presence of this flow the orientational distribution function, $\psi$ satisfies the Smoluchowski equation,
\begin{equation}
\label{smoleq}
\frac{\partial \psi}{\partial t} = \mathcal{D}_r \frac{\partial^2 \psi}{\partial \phi^2} + \dot{\gamma} \frac{\partial}{\partial \phi} \bigg(\sin^2 \phi \ \psi\bigg).
\end{equation}
The elastic stress is found from the change in the free energy according to the virtual work principle (\ref{virtualwork1}). The free energy is written, 
\begin{equation}
	F = k_B T \rho \int d\phi \ \psi \log \psi. 
\end{equation}
When the shear flow is applied, the distribution function changes and so does the free energy. The change in the distribution function is determined from the Smoluchowski equation (\ref{smoleq}). For a step shear of magnitude $\gamma$ applied over a very short time $\delta t$, the shear rate is $\dot{\gamma} = \gamma / \delta t$ and the change in the distribution function is, 
\begin{equation}
\label{distchange}
	\delta \psi = \gamma \frac{\partial}{\partial \phi} \bigg( \sin^2 \phi \ \psi \bigg),
\end{equation}
while the change in the free energy is, 
\begin{equation}
\label{FEchange1}
	\delta F = k_B T \rho \int d\phi \ \delta \psi \ \big(\log \psi + 1\big).
\end{equation}
Substituting (\ref{distchange}) into (\ref{FEchange1}) and integrating by parts we find, 
\begin{equation}
\label{FEchange2}
\delta F = -k_B T \rho \gamma \int d\phi \ \sin^2 \phi \ \frac{\partial \psi}{\partial \phi},
\end{equation}
one final integration by parts yields the elastic stress, 
\begin{equation}
\label{appstressrod}
	\sigma_{xy}  = k_B T \rho \gamma \big\langle \sin 2\phi \big\rangle.
\end{equation}
We now obtain an equation for the average $\langle \sin 2\phi \rangle$. By definition, the time derivative of this is,
\begin{equation}
	\frac{\partial}{\partial t} \big\langle \sin 2\phi \big\rangle = \int d\phi \ \sin 2\phi \ \frac{\partial \psi}{\partial t}. 
\end{equation}
Using the Smoluchowski equation and integrating by parts it is easy to show that, 
\begin{equation}
\label{avgeqn}
\frac{\partial}{\partial t} \big\langle \sin 2\phi \big\rangle = -4 \mathcal{D}_r \big\langle \sin 2\phi \big\rangle - 2\gamma \big\langle \sin^2 \phi \ \cos 2\phi \big\rangle.
\end{equation}
To calculate the linear elastic stress, the average in the final term can be replaced by its equilibrium value, $\big\langle \sin^2 \phi \ \cos 2\phi \big\rangle = -1/4$. This allows (\ref{avgeqn}) to be solved easily and using (\ref{appstressrod}) we find, 
\begin{equation}
\label{appfinstressrod}
	\sigma_{xy} = \frac{1}{2} k_B T \rho \gamma e^{-4 \mathcal{D}_r t}.
\end{equation}
So far, we have not mentioned the constraints placed on the rod which will clearly have an effect on the dynamics, altering the diffusion constant in the Smoluchowski equation.

The constraints effectively force the rod to move along its length in a channel of width $a \sim 1/(\nu L)$ and length approximately $L$. For a short time, the rod is unaffected by these constraints, since it will not have diffused far enough to be hindered by them. This is the case while it has not rotated through and angle of about $a/L \sim 1/(\nu L^2)$. If the diffusion constant without constraints is $D_r^0$ then this is true for times up to of order $\tau_{\text{int}} \sim  1/(D_r^0 \nu^2 L^4)$, as quoted in the main text. During this time period then, the stress decays according to (\ref{appfinstressrod}), but with $\mathcal{D}_r = D_r^0$.

For times longer than $\tau_{\text{int}}$, the rods orientation can only change via the reptation process. When the rod has diffused roughly $L/2$ along its length, it can disengage from its original channel, and rotate by an angle of order $a/L$. The time scale for this process is approximately $\tau \sim L^2 /D_t^0$, where $D_t^0 \sim D_r^0 L^2$ is the translational diffusion constant without constraints. Therefore the rotational diffusion coefficient estimated for the reptation process is,
\begin{equation}
	D_r(\nu) \sim \frac{a^2}{L^2 \tau} \sim \frac{a^2 D_t^0}{L^4} \sim \frac{D_r^0}{(\nu L^2)^2},
\end{equation}
as given in the main text. Hence for times longer than $\tau_{\text{int}}$, the stress decays as in (\ref{appfinstressrod}) but with $\mathcal{D}_r = D_r(\nu)$.

Putting all of these pieces together and ensuring that the stress is continuous, we recover equation (\ref{elasticresponserods}) of the main text.
\section{A more rigorous treatment of integrals}
\label{app:rigint}
Here we treat the divergent integral (\ref{FEnLDiv}) more rigorously with a small lower cutoff. This cutoff will ultimately be set by the excluded volume of the legs of the wire frame shapes. If the width of the legs is $b$, then the probability of finding a constraining point with angles less than or approximately $b/L$ must be zero due to the excluded volume effect. We define the small quantity, $\varepsilon= b/L$, and the reduced density $p = \nu L^2$ for convenience. Thus the dominant term in  (\ref{FEnLDiv}) is, 
\begin{equation}
    p^2 \int_{\varepsilon}^{\pi} d\theta \int_{\varepsilon}^{\pi} d\eta \frac{e^{-p (\theta + \eta)}}{(\theta + \eta)^2}. 
\end{equation}
The integral is taken over a square with side length $\pi$ with a smaller square with side length $\varepsilon$ taken out around the origin. To make progress without changing the scaling we approximate this region as a quarter circle of radius $\pi$ with a small quarter circle radius $\varepsilon$ removed. The integral is then written in polar co-ordinates, by defining the vectors $\textbf{r} = (\theta, \eta)$ and $\textbf{d} = (1 , 1)$ and the angle between them as $\varphi$,
\begin{equation}
    p^2\int_{\varepsilon}^{\pi} \frac{dr}{r} \int_{0}^{\pi/4} d\varphi \frac{e^{- \sqrt{2} p r \cos \varphi}}{\cos^2 \varphi}.
\end{equation}
Transforming the radial integral to the variable, $R = p r$, we have, for large $p$,
\begin{equation}
    p^2\int_{p \varepsilon}^{\infty} \frac{dR}{R} \int_{0}^{\pi/4} d\varphi \frac{e^{- \sqrt{2} R \cos \varphi}}{\cos^2 \varphi}.
\end{equation}
The $R$ integral can then be evaluated as an incomplete gamma function\cite{Abramowitz1964HandbookTables}, 
\begin{equation}
    p^2 \int_{0}^{\pi/4} \frac{d\varphi}{\cos^2 \varphi} \ \Gamma(0,\sqrt{2} p \varepsilon \cos\varphi).
\end{equation}
Since we are taking the density to be in the range, $1/L^2 \ll \nu \ll 1/V_{exc}$, we may take the limit $\nu \varepsilon \to 0$, giving
\begin{equation}
    -p^2 \int_{0}^{\pi/4} \frac{d\varphi}{\cos^2 \varphi} \ \Big[\gamma_{em} + \log(\sqrt{2} p \varepsilon \cos \varphi)\Big],
\end{equation}
where $\gamma_{em}$ is the Euler-Mascheroni constant. The integral can now be taken exactly using integration by parts for the second term,
\begin{equation}
\begin{split}
    &-p^2 \Big[\tan \varphi \bigg(\gamma_{em} + 1 + \log(\sqrt{2} p \varepsilon \cos \varphi)\bigg) - \varphi \Big] \Big\lvert_{0}^{\pi/4} \\
    & = p^2 \big(|\log p \varepsilon|  - C \big),
\end{split}
\end{equation}
where we have used $p \varepsilon \ll 1$, to make the sign more obvious and defined the positive, order unity constant, $C$. 

Hence we find only logarithmic corrections to the $\nu^2$ scaling discussed in the main text. This does not change the qualitative features of our results. 
\section{Evaluation of Bending Probability Integrals} \label{app:bendint}
To get the contribution to the bending probability from region (I), see (\ref{regions}), we need to evaluate the integral,
\begin{equation}
\frac{1}{4}\int_{0}^{\nu L^2 \theta_0(\phi)} dy_0 \int_{0}^{\nu L^2 \eta_1(y_0/ \nu L^2)} dx_1 e^{-(x_1 + y_0)/2}.
\end{equation}
Where the functions, $\theta_0(\phi)$ and $\eta_1(y/\nu L^2)$ are defined in equations (\ref{theta0phi}) and (\ref{eta1oftheta0}) of the main text respectively. Taking the $x_1$ integral gives, 
\begin{equation}
\label{PB5}
\frac{1}{2}\int_{0}^{\nu L^2 \theta_0(\phi)} dy_0 \bigg(e^{-x_1/2} - e^{x_1/2-\nu L^2 \eta_1(y_0/ \nu L^2)/2}\bigg).
\end{equation}
Integrating the first term in the integrand is straightforward whereas the second integral is schematically,
\begin{equation}
	\int_0^{c \gamma + d \gamma^2} dx \ e^{\nu L^2 \gamma a(x/\nu L^2) + \nu L^2 \gamma^2 b(x/\nu L^2)}.
\end{equation}
The integrand can be expanded in powers of $x/\nu L^2$, because over the whole integration range $x/\nu L^2 \sim \mathcal{O}(\gamma)$,
\begin{equation}
		\int_0^{c \gamma + d \gamma^2} dx \ e^{\nu L^2 \gamma a(0) + \nu L^2 \gamma^2 b(0)}\Big[1 + x \big(\gamma a'(0) + \gamma^2 b'(0)\big)\Big],
\end{equation}
clearly the integral of the second term in the square brackets will be $\mathcal{O}(\gamma^3)$, and so only the first term contributes to second order. Therefore and the result of the integral (\ref{PB5}) is, 
\begin{equation}
1 - e^{-\nu L^2 \theta_0(\phi)/2} -\frac{1}{2}\nu L^2 \theta_0(\phi) e^{- \nu L^2 \eta_0(0)/2},
\end{equation}
as given in equation (\ref{bendprobint}) of the main text. 
\section*{Data Availability}
The data that support the findings of this study are available from the corresponding author upon reasonable request.
\bibliography{references}
\end{document}